\documentclass{iopart} 
\usepackage{iopams}  
\usepackage{graphicx,graphics}	
\graphicspath{{images/},{figures/}{Eduardo2/}}
\usepackage[utf8]{inputenc}
 \expandafter\let\csname equation*\endcsname\relax 
 \expandafter\let\csname endequation*\endcsname\relax
\expandafter\let\csname equation*\endcsname\relax 
\expandafter\let\csname endequation*\endcsname\relax
\usepackage{amsmath,amssymb,bbm} 
\usepackage{color}

\def\notext#1{}

\newcommand{\mcH}{\mathcal{H}}	
\newcommand{\identity}{\mathbbm{1}}
\newcommand{\1}{\identity}

\renewcommand{\>}{\rangle}
\def\1{{\mathchoice{\rm 1\mskip-4mu l}{\rm 1\mskip-4mu l}{\rm 1\mskip-4.5mu l}{\rm 1\mskip-5mu l} }}

\newcommand{\KI}{\mathrm{KI}}

\usepackage{hyperref}

\providecommand{\openone}{\leavevmode\hbox{\small1\kern-3.8pt\normalsize1}}

\newcommand{\ifunamint}{\address{$^3$Instituto de Física, Universidad Nacional Aut\'onoma de M\'exico, 01000 M\'exico D.F., Mexico}}
\newcommand{\facunam}{\address{$^2$Facultad de Ciencias, Universidad Nacional Autónoma de México, 01000 M\'exico D.F., Mexico}}
\newcommand{\cic}{\address{$^4$Centro Internacional de Ciencias A. C., Avenida Universidad s/n, 62131 Cuernavaca, Morelos, Mexico}}
\newcommand{\icf}{\address{$^1$Instituto de Ciencias Físicas, Universidad Nacional Aut\'onoma de M\'exico, 62210 Cuernavaca, Morelos, Mexico}}

\begin{document} 
\title[Stabilizing coherence with nested environments]{Stabilizing coherence with nested environments: a numerical study using kicked Ising models} 
\author{
C González-Gutiérrez$^1$, 
E Villaseñor$^2$, C Pineda$^3$ and 
T H Seligman$^{1,4}$}
\icf
\facunam
\ifunamint
\cic

\date{\today}%
\begin{abstract}
We study a tripartite system of coupled spins, where a first set of one or two
spins is our central system which is coupled to another set considered, the
near environment, in turn coupled to the third set, the far environment.  The
dynamics considered are those of a generalized kicked spin chain in the regime
of quantum chaotic dynamics.  This allows to test recent results that suggest
that the presence of a far environment, coupled to the near environment, slows
decoherence of the central system.  
After an extensive numerical study, we
confirm previous results for extreme values and special cases. 
In particular, under a wide variety of circumstances an increasingly large
coupling between near and far environment, slows decoherence, as measured by
purity, and protects internal entanglement.
\end{abstract}
\section{Introduction} 

Conserving quantum coherence for long times is a challenge for quantum
information processing and quantum computation implementations.  Mechanisms for
quantum coherence protection like decoherence free subspaces and dynamical
decoupling are very popular nowadays~\cite{lidar14}. In many situations the
immediate environment ({\it near environment}) surrounding a quantum system can be
well controlled in order to isolate the system for unwanted external interactions.
What about the rest of the universe, i.e., the environment that is not immediately 
accessible to the system ({\it far environment})? 
Can it affect the decoherence process of the
system due to the interaction with the near environment?  More specifically,
can it slow down the decoherence (or equivalently, protect the coherence) in
the system?  In this work we investigate these questions in a particular situation
in which both environments (near and far) are modelled by a quantum kicked spin
system.

Recently there has been an increasing awareness, that nested environments can
actually be used to control the coherence of a central system~\cite{Zanardi14,moreno2015}.
While it seems that such effects for some time were known to occur in the context of the 
Jaynes-Cummings model with leaky
cavities~\cite{nemesprivate,TorresandSeligman}, i.e., in  the setting of
Haroche's
celebrated experiment~\cite{Raimond97,Brune96}, there has been increasing
interest in this matter recently~\cite{Zanardi14,moreno2015,moreno13,Gonzalez14}.  
In the present paper we plan to broaden the work presented in the thesis of one of
us~\cite{Gonzalez14}. We shall  study a situation where a central system of
one or two quits is coupled to a near environment which, in turn, is coupled to
a far environment. Direct  coupling of the system to a far environment is
assumed small enough to be neglected. We furthermore assume that the coupling
between the central system and the near environment is weak, in order to obtain
a situation, that is meaningful for quantum information processes. To fix ideas
we may think of an ion quantum computer as developed in
Innsbruck~\cite{Haffner2008}, with very stable ions, that suffer some
decoherence because of the apparatus, mainly in the control gates, and the
additional degrees of freedom brought into play by this step in turn have
further decay modes or fluctuation sources, not affecting the central system
directly in a significant way. This, we can think of as a far environment.
Considering previous work, the aim of this paper can be formulated more
precisely: We wish to confirm as far as possible previous work, which to a
large concentrates on small couplings and very strong couplings to the far
environment.  Yet the main purpose is to explore in more detail the
intermediate zone with a structured dynamic model, that allows for the
necessary flexibility and large scale calculations. This will include the use
of dephasing for the coupling between central system and near environment for
two reasons: First because it relates to a generalized concept of fidelity
discussed in~\cite{gorin2015generalized} and second because dephasing clearly
separates our results form similar ones seen for the energy
transfer~\cite{nemesprivate,TorresandSeligman,Bosco11}. 

We plan to use the kicked Ising model~\cite{prosen02} as a paradigmatic example
to construct such a situation, because of its great flexibility, as well as for
its numerical
efficiency~\cite{prosen02,pineda14,prosen-seligman02,pizorn08,pizorn07,pinedaPRE06}.
While the original model was concerned with nearest neighbours coupled spin
chains, more general situations have been studied. The generalization used
in~\cite{pinedaPRA06} is relevant for this work in order to deal with more
complicated configurations necessary to handle the decoherence process.
 
The program for this paper is to use this very flexible and efficient model to
cover a wide range of situations for the couplings.  This implies in the first
place the coupling strength between central system and near environment
($\lambda$) as well as the one between near and far environment ($\gamma$).
Furthermore, the configuration and numbers of the connections between the three
subsystems  can be varied and will play a significant role.  For the central
system on the other hand we take the simplest options, namely a single qubit or
two non-interacting qubits, and we focus our attention on the evolution of
purity and concurrence.

The paper is organized as follows:
In~\sref{sec:model}  we introduce the model and
the
decoherence and entanglement measures used along the paper are introduced. 
In~\sref{onequbit} we present the results for the case of one qubit as a
central system.  Here, we confirm the basic hypothesis that for weak
coupling of the central system to the environment increasing the coupling 
of the latter to the far environment will slow down
decoherence.
Moreover, we shall study in detail the dependence of such behaviour on 
the parameters of the model.
The analysis for the case of two non-interacting qubits
is presented in~\sref{twoqubits}, where one can analyse a similar
effect on the internal
entanglement of the central system.
We finish the paper in~\sref{sec:conclusions} with the conclusions.

\section{The model}  
\label{sec:model}
\subsection{The kicked Ising system}  
\label{sec:kimodel} 
We shall follow the spirit of~\cite{pinedaPRA06} and establish a kicked Ising
model (KI) as a tool to analyse numerically the time evolution of coupled quantum
systems and their entanglement. The model was developed by T. Prosen~
\cite{prosen02}, who showed that this system is amenable to large scale
computation and indeed allows that treatment of large numbers of qubits
efficiently including packages for the use of GPUs~\cite{pineda14}.

We can think of our model as a graph of $N$ qubits (spin $1/2$
particles) connected by $N(N-1)/2$ Ising connections of any
strength~\cite{pinedaPRA06} and a single qubit term, which will be periodically
kicked with a magnetic field.  This is
described by the Hamiltonian 
\begin{equation}
\label{eq:KI}
 H_{\KI}=\sum_{j>k}^N 
    J_{j,k}\sigma^j_{z}\sigma^{k}_{z}+K(t)\sum_{j}^N\vec{b}_j\cdot\vec{\sigma}^{j},
\end{equation}
with $\sigma^{j}_{x,y,z}$ being the Pauli matrices of spin $j$ and
$\vec\sigma^j=\left(\sigma^j_{x},\sigma^j_y,\sigma^j_z\right)$.
The time-dependent function $K(t)=\sum_{n\in\mathbb{Z}}\delta(t-n)$ is a train
of Dirac-delta functions of period one. The evolution of the KI is described by
means of the Floquet operator $U_\KI$, which is the evolution operator for one
period of time.  During the free evolution, the system evolves with the unitary
propagator corresponding to the Ising interaction
\begin{equation}\label{ising}
U_{\text{Ising}}
  =\exp\left(-\text{i}\sum_{j>k}^N J_{j,k} \sigma^j_{z}\sigma^{k}_{z}\right)
  =\prod_{j>k}^N\exp\left( -\text{i}J_{j,k}\sigma^j_{z}\sigma^{k}_{z}\right),
\end{equation}
and the effect of the magnetic kick is obtained with the one-qubit operator
\begin{equation}
 U_{\text{Kick}}
   =\exp\left(-\text{i}\sum_{j=1}^N\vec{b}_j\cdot\vec{\sigma}^{j}\right)
   =\prod_{j=1}^N\exp\left(-\text{i}\vec{b}_j\cdot\vec{\sigma}^{j}\right),
\end{equation}
where we take the units such that $\hbar=1$. Therefore, the Floquet operator
for one period of time is
\begin{equation}
U_{\KI}= U_{\text{Kick}} U_{\text{Ising}}.
\end{equation}

Originally, the KI  consists of a ring of $N$ spins-$1/2$ (or qubits)
interacting via homogeneous nearest-neighbour dimensionless Ising coupling $J$ and are
periodically driven by a uniform dimensionless magnetic field $\vec{b}$.  
This system is trivially integrable in the case of magnetic field parallel to the
Ising axis. It was showed that integrability still exists if the magnetic field
is orthogonal to the Ising interaction~\cite{Prosen00,arul05}. 
The two-dimensional version of the KI was studied recently using an efficient time
evolution for the Floquet operator~\cite{pineda14}, where evidence of the
chaoticity of the transverse case is presented, in the form of statistics of
the eigenphases of the corresponding Floquet operator.

Another important family of models is obtained when instead of having a fixed,
and small, number of parameters, one allows some of these parameters to be
random (though static).  In particular, the random two-body interaction
Hamiltonians used in~\cite{pizorn08} can readily be included in our framework, by
considering nearest neighbour interactions with random strength plus a single
particle term (in our case not random,  for simplicity). 
\subsection{Two nested environments in the KI}  
\begin{figure} 
 \centering
 \includegraphics[scale=1.3]{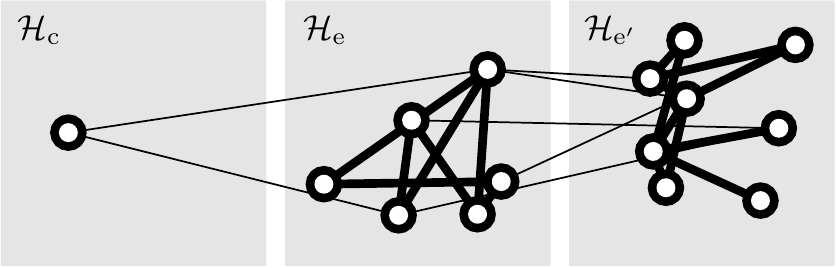}
 \caption{Example for a tripartite configuration of qubits systems. 
 Each of the Hilbert spaces composing the total space
 \eref{eq:hilbert:space:structure}, is displayed as a shaded area. The open 
 circles represent the different qubits, thick lines represent strong
 interactions, and thin ones weak interaction. 
}
\label{fig:general:ceep}
\end{figure} 
Regarding the qubits as nodes, and non-zero Ising couplings as connections, we
can regard such system as a graph.  While the analysis of an arbitrary
graph will be of interest in the light of recent results for complete quantum
graphs in the context of random matrix theory (RMT)~\cite{Weidenmuller14}, 
such a generalization will not be the subject of the present paper.

If we further allow to have two different types of connections (in our case
strong an weak), we can have a collection of connected graphs.  We will thin
out or dampen connections to obtain the formal structure of a central system
and nested environments we need.  To be more precise, we shall work in a
Hilbert space structured as 
\begin{equation}
\mcH = \mcH_{\text{c}}\otimes\mcH_{\e}\otimes\mcH_{\e'},
\label{eq:hilbert:space:structure}
\end{equation}
where each the Hilbert spaces denoting central system, near environment and far
environment, respectively. With respect to the spin environments, we shall
consider that within each Hilbert space, there is strong coupling, and from the
space of the central system to the near environment weak coupling is assumed,
see~\fref{fig:general:ceep}. We want to investigate the effect  of different
coupling regimes from the space of near environment to the far environment on
the weakly coupled central system.

We shall limit the central system $\mcH_{\text{c}}$ to simple cases of one or
two qubits. The former is obviously the simplest non-trivial system we can have
and the latter is the building block for quantum information systems as quantum
gates involving pairs of qubits are sufficient to represent a universal quantum
computer~\cite{nielsen2010}.  Furthermore, as we are not interested in the
operation of quantum gates, we shall actually turn off the interaction between
these to qubits, i.e., treat a quantum memory, which for weak decoherence can be
understood entirely in terms of single-qubit decoherence~\cite{Gorin07}.  We
shall separate  the rest of the system into two sets of qubits forming $\mcH_{\e}$ and $\mcH_{\e'}$, respectively. 
For example, these qubits can be organized
in an open chain where $\mcH_{\e}$ will be connected 
to the central system at one spin and to the far environment by connecting a spin of this environment to another (or the same) spin of the central system.
 $\mcH_{\e'}$ will have no connections to $\mcH_\text{c}$.  Within each of these strings we shall
consider nearest-neighbour interaction.  We will also present some results for
additional couplings within each subset, which for quantum chaotically
situations result rather irrelevant, as may be expected following universality
arguments.  For the coupling between the two subsets we will vary from  a
single coupling to many pairs interacting. 

We can reorganize the Hamiltonian~\eref{eq:KI}, and write it as
\begin{equation}
H=H_{0}+\lambda V_{\text{ce}}+\gamma V_{\text{ee}'},
\end{equation}
with 
\begin{equation}\label{general_model}
 H_{0}=H_{\text{c}} +  H_{\e} +\ H_{\e'},
 \end{equation}
where the indices denote the Hilbert spaces in which the operators act
non-trivially.  $\lambda$ and $\gamma$ are real non negative coupling
parameters between central system and near environment and between near
environment and far environment, respectively.  The Hamiltonian $H_{0}$
represents the internal dynamics of the central system, the near environment
and the far environment.  The operators $V_{\text{ce}}$ and $V_{\text{ee}'}$
represent the interaction between each shaded area
in~\fref{fig:general:ceep}. To be more precise, we can label the set of spins
belonging to $\mcH_{\kappa}$ as $S_{\kappa}$ with $\kappa=\text{c}$, $\e$ or $\e'$. Then, 
\begin{equation}
H_\kappa= J \sum_{j>k \in S_\kappa} I_{j,k}^{(\kappa)}\sigma^j_z\sigma^k_z
  +K(t)\sum_{j\in S_\kappa}\vec{b}_\kappa\cdot\vec{\sigma}^j,
\label{eq:puto:cafe}
\end{equation}
and $I_{j,k}^{(\kappa)}$ is a matrix of
zeros and ones containing the configuration of the subsystem. 
The {\it open chain} system where $I_{j,k}^{(\kappa)}
= \delta_{k+1,k}$ will be of particular interest.
The interaction between central system and near environment and the one between
near and far environment are given by the Ising terms
\begin{equation}
V_{\text{ce}}=\sum_{j \in S_\text{c}, k\in S_{\e} }I^{(1)}_{jk}
                \sigma^j_z\sigma^k_z,\quad
V_{\text{ee'}}=\sum_{j \in S_\e, k\in S_{\e'} }I^{(2)}_{jk} \sigma^j_z\sigma^k_z
\label{eq:interacciones}
\end{equation}
respectively, where in this case $I^{(1,2)}_{jk}$ are other matrices with zeros or ones,
containing the particular configurations of the interactions.  
 Notice that the propagator \eref{ising} is periodic (up to a global phase) in
$J_{j,k}$, 
as $\exp[ \text{i} (J_{jk} + \pi) \sigma^j_{z}\sigma^{k}_{z} ] = - 
\exp[ \text{i} J_{jk}  \sigma^j_{z}\sigma^{k}_{z} ]$,
so the behaviour of all observables
will also be periodic in $\gamma$ and $\lambda$ with period $\pi$. 
We have also observed a symmetry of the channel induced in the 
central system with respect to sign changes both in $\gamma$ and $\lambda$, 
so it will suffice to study, with respect to both parameters, the
interval $[0,\pi/2]$. We however will show one example for a full period
of $\gamma$ to illustrate the effect.

 Taking into account that both the free and the kicked part can be decomposed 
in a simple multiplication of one and two qubit operations, we can 
see that this model can be numerically evolved efficiently. The memory requirements
are
set by the size of the state to be evolved ($2^n$, where $n$ is the total number 
of qubits), and the speed of the algorithm is also linear with respect to the
size of the Hilbert space, for each time step to be evolved. We have used
efficient tools~\cite{pineda14} developed for GPUs (graphics processing unit)
to do our numerical
experiments.

The dynamics of the central system is obtained tracing over both environments,
which leads to the reduced density operator
\begin{equation}
 \rho_{\text{c}}(t)=\tr_{\e,\e'}[\rho(t)],
\label{eq:evolution:rho}
\end{equation}
where the total density operator evolves unitarily as
$\rho(t)=U_{\text{KI}}(t)\rho(0)U^\dagger_{\text{KI}}(t)$.  We  also assume 
the absence of initial quantum correlations between subsystems, i.e., the initial state
for the total system is separable
\begin{equation}
\rho(0)=\rho_{\text{c}}\otimes\rho_{\e}\otimes\rho_{\e'}.
\end{equation}
In all numerical results we will present, two kinds of initial pure states for
the central system are used.  For a one qubit central system, $\rho_{c}(0)$ will
be taken as an eigenstate of the operator
$\sigma_{x}$, while for a two qubit central system we shall use the Bell state 
$(|00\>+|11\>)/\sqrt2$,  
which gives us the opportunity to study the evolution of internal
entanglement within the central system. In order to emulate a high-temperature spin
bath, the initial state of the environments is chosen as a product of 
two random pure states, one for each environment. 
\subsection{Quantifying decoherence and entanglement} 
In order to measure the loss of coherence in our central system we use the
purity of a density operator defined as
\begin{equation}\label{purity}
 P[\rho]=\tr{\rho^2}.
\end{equation}
This quantity varies from $1/\dim(\rho)$ for the totally mixed state to 
a vale of one for pure states. If $\rho$ is the partial trace of a pure state, 
as in \eref{eq:evolution:rho}, it measures the entanglement 
between the two subsystems implied in the partial trace operation. 
Purity is just one of a large number of convex functions that can describe
decoherence. Its main advantage is the simple analytic structure which allows
to compute it without previous diagonalization of the density matrix. Another
commonly used convex function is the von Neumann entropy. 
Partial orders can be obtained using all or complete sets of
positive functions \cite{Uhlmann,ru1,ru2}. Any of these convex functions
reveal, in general, different aspects of decoherence. In fact, for a single qubit
they are all equivalent and for larger systems,  near pure states, 
they also tend to be equivalent.

On the other hand, a good measure to quantify the entanglement shared between two qubits is the so called
concurrence, which is defined for a general mixed state as~\cite{Wootters98}
 \begin{equation}\label{concurrencia}
   C(\rho)=\max\{0,\tilde\lambda_1-\tilde\lambda_2-\tilde\lambda_3-\tilde\lambda_4\},
 \end{equation}
where $\tilde\lambda_{i}$ are the square roots of the eigenvalues of
$\rho\tilde\rho$ in decreasing order. The operator $\tilde\rho$ is the result
of applying a ``spin flip'' operation on $\rho$, i.e.,
$\tilde\rho=(\sigma_{y}\otimes\sigma_{y})\rho^{*}(\sigma_{y}\otimes\sigma_{y})$
and the complex conjugate is taken in the computational basis of two qubits.
 
Decoherence of one and two qubits measured by the purity and the entanglement
of the former trough the concurrence in an environment described by a kick
Ising chain is the main study of~\cite{pinedaPRA06}.
\section{One-qubit in a nested environment} 
\label{onequbit}
We shall first explore the effect of nested environments with the simplest
central system possible, namely a single qubit. 
The coupling between the central system and the near environment will be 
chosen to be weak, while the coupling of the near environment to the far environment
will range  from weak to strong but always stronger that the former. As in previous works
we neglect any coupling between the central system and the far environment, 
but we will test this assumption in a typical situation. Throughout the paper,
we shall choose the dimensions as large as we could for both  the near and far
environments
without having excessive computation times. We also set the parameters of the model 
in a regime where the dynamics of the chains
are in the quantum chaotic regime~\cite{prosen02}, so as to mimic universal
results~\cite{haakebook}. Specifically, we set
$\dim(\mcH_{\text{e}})=2^{6}$, $\dim(\mcH_{\text{e}'})=2^{10}$,
$J=1$ and $\vec{b}_\e = \vec{b}_{\e'}=(1,0,1)$ in Eq.~\eqref{eq:puto:cafe}.
This implies that the kicked magnetic field has an angle of $\pi/4$ with
respect to 
the Ising coupling, and the field strength is chose appropriately. 
The coupling between the central system and the near environment, 
shall be fixed to $\lambda=0.01$ unless otherwise 
stated. Similarly, the size of the near and far environment
shall be fixed as in~\fref{puritytime}, except when we analyse the
effect of the dimensions in the environments.
Finally we choose for most of this section a dephasing coupling between near
and far environment (i.e. $\vec{b}_\text{c} = 0$). This is inspired by the
observation~\cite{gorin2015generalized} that it will lead to a measurable
fidelity amplitude for the open near environment using the central qubit as a
probe as in the original quantum optical proposal~\cite{Zoller98,
Schleich05,gorin2004} for the measurement of fidelity decay, but now applied to
an open system, which results by adding the far environment.  We obtain the
dephasing case by dropping the kicked magnetic field for the central qubit,
whose Hamiltonian then commutes with the Ising coupling to the near
environment.  Another advantage of the dephasing case is that it involves no
energy transport and thus clearly distinguishes the decoherence behaviour from
results for energy transport.  Towards the end of this section we shall lift
this restriction.  

Our model depends on the configuration of the connections used within each
environment, the configuration of the connections between environments and the
number of such connections as well as on their strength. The main object of
this paper is to study the behaviour of purity on the aforementioned details.
The exploration cannot be exhaustive; rather we tested typical changes and we
test in each of these cases the behaviour of decoherence.

The configuration we first consider is characterized by including a single
connection between both the central system and the near environment (an open
chain), and between the latter and the far environment (also an open chain),
see fig~\ref{puritytime}.  The decay of the purity of the central system as a
function of time for fixed $\lambda$ and for different values of $\gamma$ is
shown in~\fref{puritytime}.  This picture shows the main feature we wish to
discuss in this paper: {\it larger values of $\gamma$ lead to slower
decoherence in the central system}.  At this point we repeat the same
calculation as in \fref{puritytime}, but adding a weak Ising coupling of $0.01
\lambda$ between one qubit in the far environment and the central qubit.  As
expected, the resulting plot is indistinguishable, from the aforementioned
one, and we thus do not present it.  

We shall now proceed to look at who details of our model affect this result.

First we shall explore how the configuration of the internal connections in
each of the environments with unchanged connections between the subsystems
impact the results.  In \fref{puritygamma} we show the configurations, shown in
the upper part, and plot the purity reached at a given time as a function of
$\gamma$ for each of these configurations.  We limit $\gamma$ to the range
$[-\pi/2 ,\pi/2]$ as the function must be periodic with period $\pi$ (see
\sref{sec:kimodel}).  The figure shows that adding more internal connections in
the environments does not qualitatively change the  behaviour of purity.  As we
allow $\gamma$ to become larger we see that the tendency is reversed due to the
reflection symmetry discussed in \sref{sec:kimodel}. Note that we see a peak
near $\gamma=0$. This is no contradiction, but rather confirms that we need $
\gamma>\lambda$ to assure that we see our effect.

\begin{figure} 
 \centering
 \includegraphics[scale=0.5]{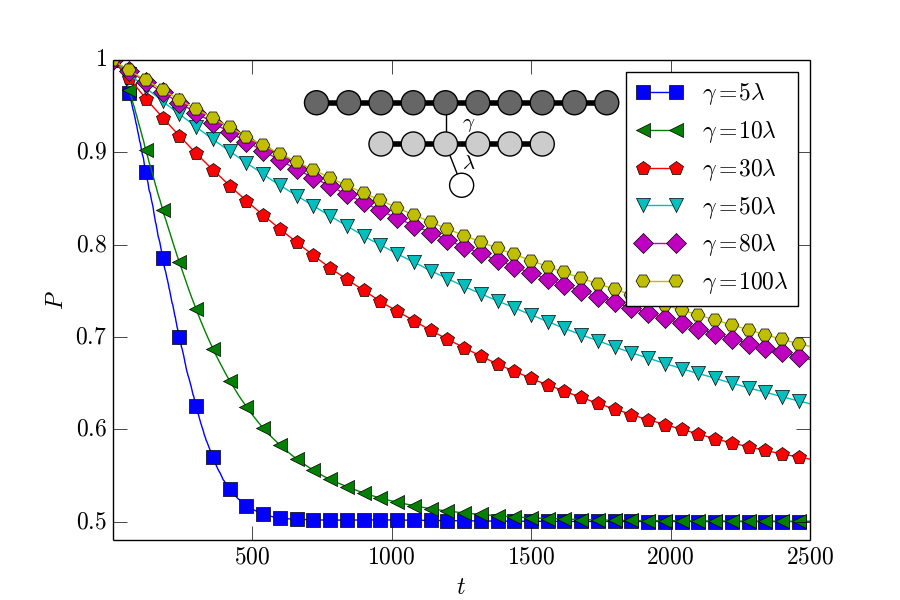}
 \caption{Purity decay of the central system (open circle)
 for $\lambda=0.01$ and different
 values of $\gamma$, for the configuration illustrated in the upper part
 of this figure.  Larger values of the coupling between the near
 (grey circles)
 and far (black circles)
 environments induce less decoherence of the central system.
 }
 \label{puritytime}
\end{figure} 

\begin{figure} 
 \centering
 \includegraphics[scale=0.4]{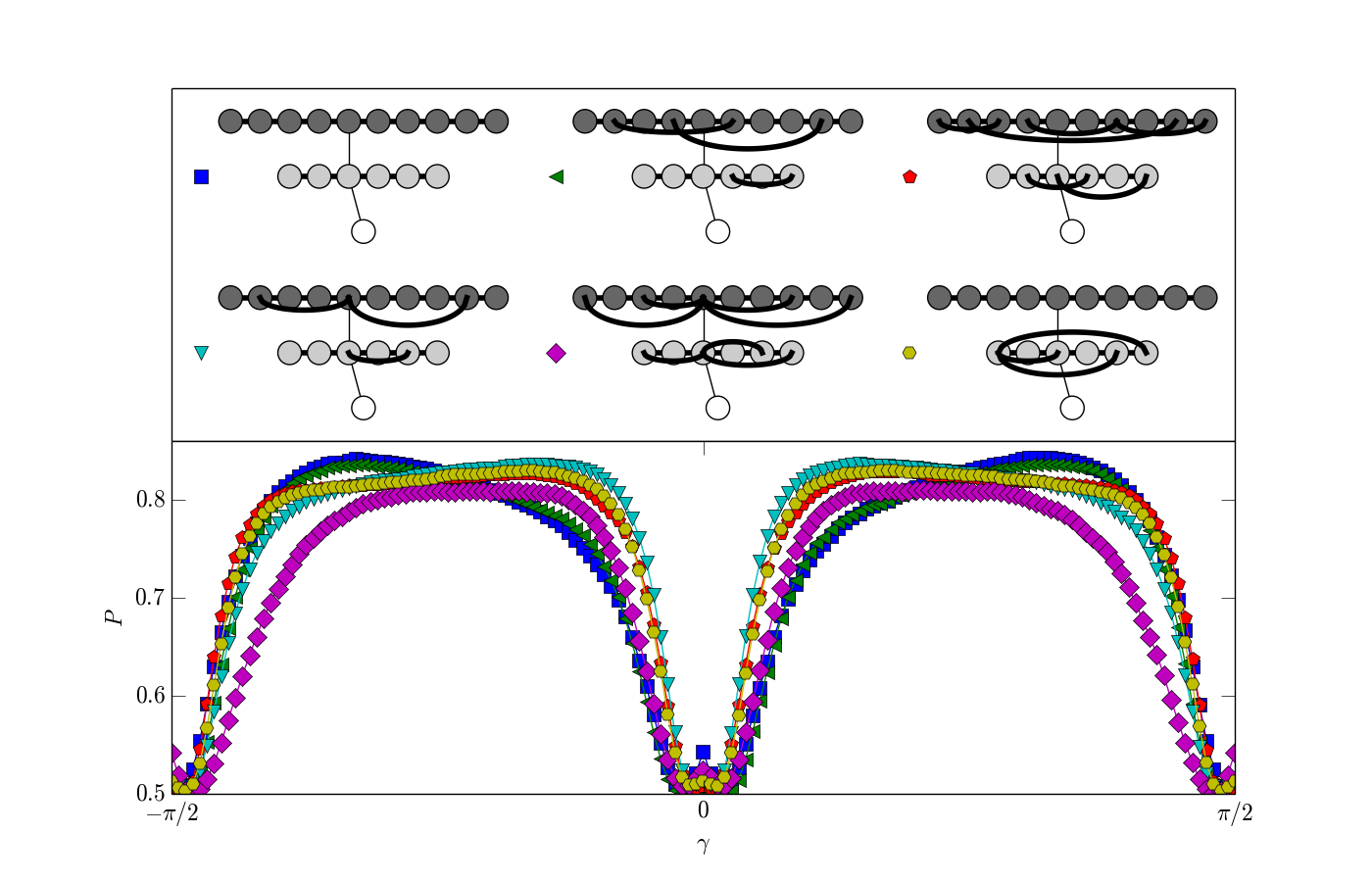}
 \caption{Purity as a function of $\gamma$ for a fixed time $t=1000$. The
 dimensions and parameters are the same as in \fref{puritytime}. Six cases are
 shown, the first of the simple direct connection and additional five ones where
 additional two body connections are made within the environments as the
 diagrams show. Results show a plateau where a maximum purity value is reached.
 We see how the extra connections within the environments appear to have little
 effect in the overall phenomena. }
 \label{puritygamma}
\end{figure} 

Next we test to what extent more connections between the environments have an
effect in the decoherence process.  We have found an interesting result:
additional connections between the environments appears to have little effect
over the decay of purity,
except for the implicit strengthening of the coupling. This can be compensated
approximatively by scaling $\gamma$ with 
$1/\sqrt\nu$ of $\gamma$, i.e., the decays are very similar for a number $\nu$
of connections chosen randomly, if simultaneously $\gamma$ is replaced by
$\gamma^{\prime}=\gamma/\sqrt\nu$, as we show in \fref{gammascaled}.
\begin{figure} 
 \centering
 \includegraphics[scale=0.5]{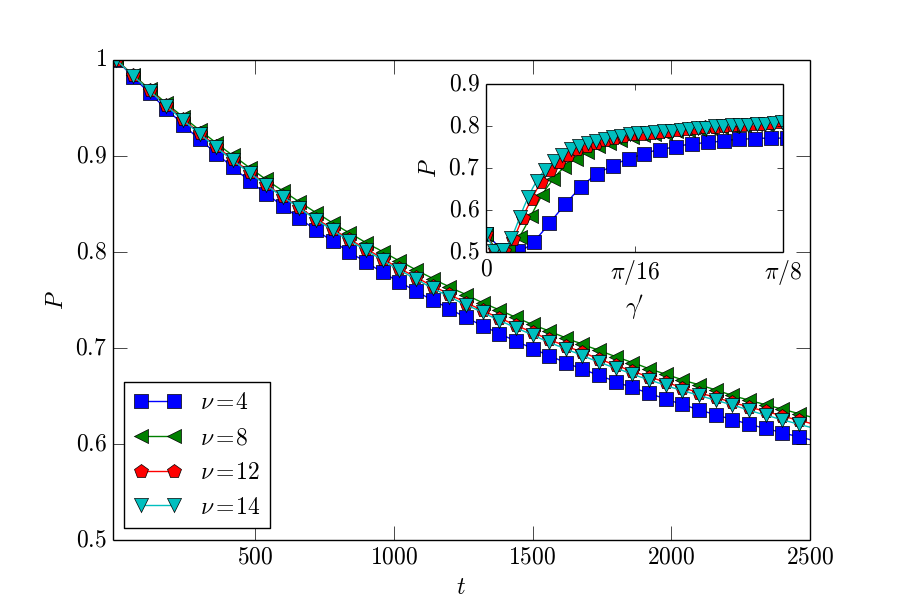}
 \caption{Purity decay over time for different number of connections. In each
 case the parameter $\gamma^\prime=\gamma/\sqrt{\nu}$ is used. The inset
 shows purity, for $t=1000$, as a function of $\gamma^\prime$. We see how using
 the value $\gamma^\prime$, for sufficiently large values of $\gamma'$ there is
 no gain in increasing the number of
 connections as all the cases show basically the same behaviour.}
 \label{gammascaled}
\end{figure} 
An interesting phenomena appears as we take the connections between
environments randomized, while conserving their total number $\nu$. We
found that the configuration of the connections appears to have a definite effect,
specially for large values of $\gamma$. In \fref{connections}
we show how different topologies affect the decay in different ways.
One can notice that, in this case, different topologies cluster around three
different behaviours, one with a flat plateau, with respect to $\gamma$, other
with a slight, but
noticeable maximum around $\gamma=\pi/4$, and finally another one with a skew 
behaviour. This is not so for other configuration of the internal connections, 
as can be seen in \fref{puritygamma}, but it is also true for more connections
between the environments, as can be seen in \fref{connections2}.
We were not successful in determining the
pattern that lead to one 
or the other behaviour. 
Additionally \fref{connections2} shows how for different numbers $\nu$ the
configuration of the connections is what has an impact on purity decay, and not the
number of connections, as long as they are 
compensated by proper rescaling.

\begin{figure} 
 \centering
 \includegraphics[scale=0.4]{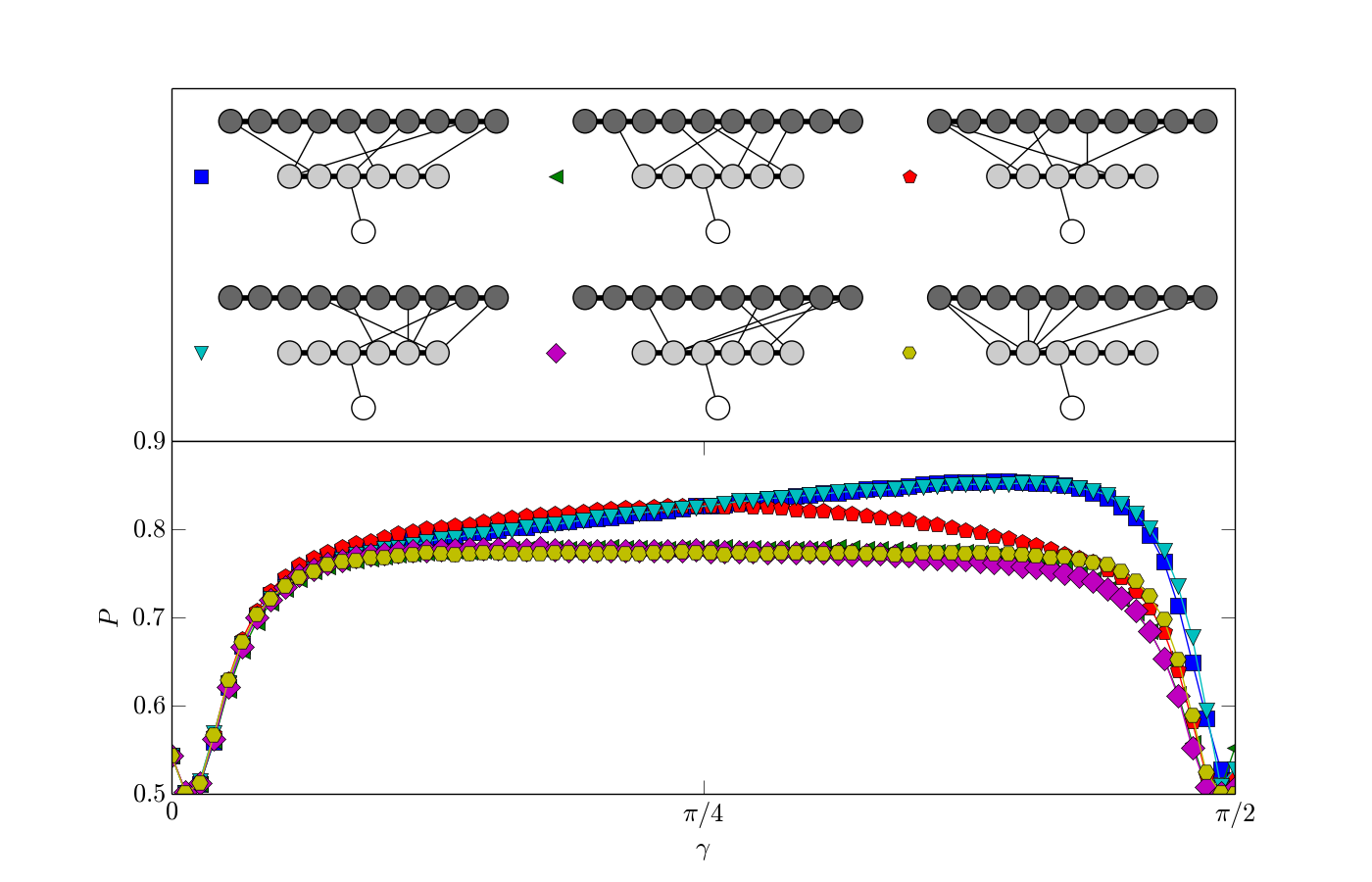}
 \caption{Purity as a function of $\gamma$ for a fixed number of connections
 $\nu=6$. In each case a different configuration of connections is used between
 environments, each shown in as a diagram. Results show the existence of three
 qualitatively different behaviours for the purity depending strongly on the
 configuration. The first one corresponds to a flat plateau, the second one to an
 increasing but symmetric in $\pi/4$ shape and the last one to a increasing non
 symmetric shape. }
 \label{connections}
\end{figure} 

\begin{figure} 
 \centering
 \includegraphics[scale=0.4]{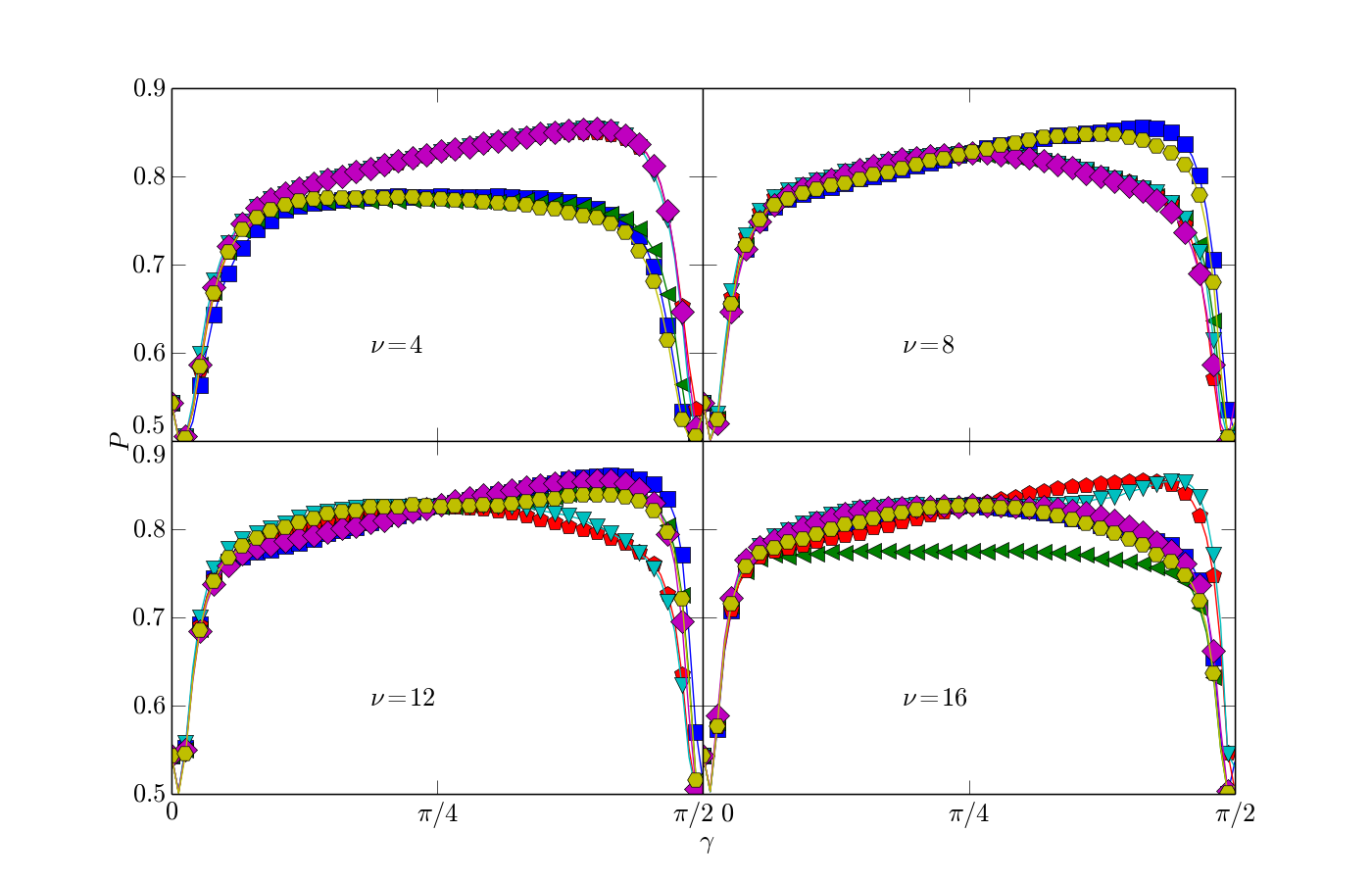}
 \caption{Purity as a function of $\gamma$ for four different sets of
 topologies each with a fixed number of connections $\nu$, each curve represents a
 different randomly selected configuration of connections between environments.
 We see how the three behaviours are present in the different cases, showing
 that increasing the number of connections has little effect in the
 predominance of a specific behaviour.}
 \label{connections2}
\end{figure} 

We next study the effect of the dimensions of the environments on purity decay.
In~\fref{sizes} we show for a single direct connection [see \fref{puritytime}]
how, as the dimension increases, 
the maxima for purity becomes ever broader but the increase is slow.
Maybe this is even hinting the
possibility of this happening for all non zero values of $\gamma$ in the case
where the dimensions of the environments go to infinite,
but we shall later see, that there is strong indication that we must also relax the 
dephasing condition in order to reach that limit.  
\begin{figure} 
 \centering
 \includegraphics[scale=0.4]{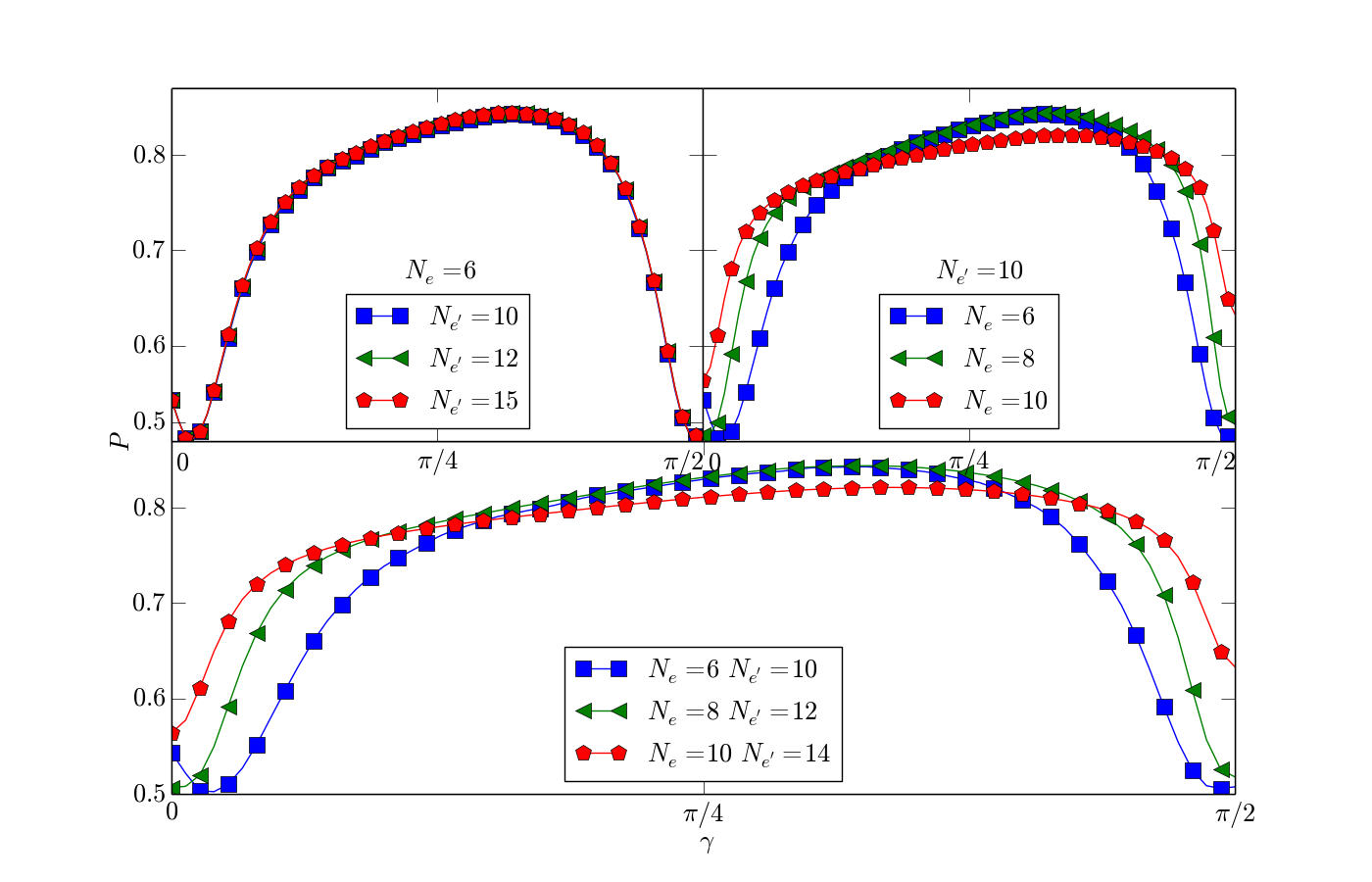}
 \caption{Purity as a function of $\gamma$ for the simple direct connection
 configuration for different sizes of environments, where
 ${\rm dim(H_e)}=2^{N_e}$ and ${\rm dim(H_{e^\prime})}=2^{N_{e^\prime}}$. We
 see how increasing the environment sizes does not increases 
 the effect of the phenomena, as it only makes it more persistent over the parameters.}
 \label{sizes}
\end{figure} 

As a final test we have to look at the $\lambda$ dependence of the effect; as
the interaction of the central system with the near environment is unavoidable,
we have to check how small it must be to actually obtain an improvement of
coherence if we increase the interaction of the near environment with the far
environment. For this purpose we show the purity as a function of $\lambda$ in
\fref{logpurity}.

\begin{figure} 
 \centering
 \includegraphics[scale=0.5]{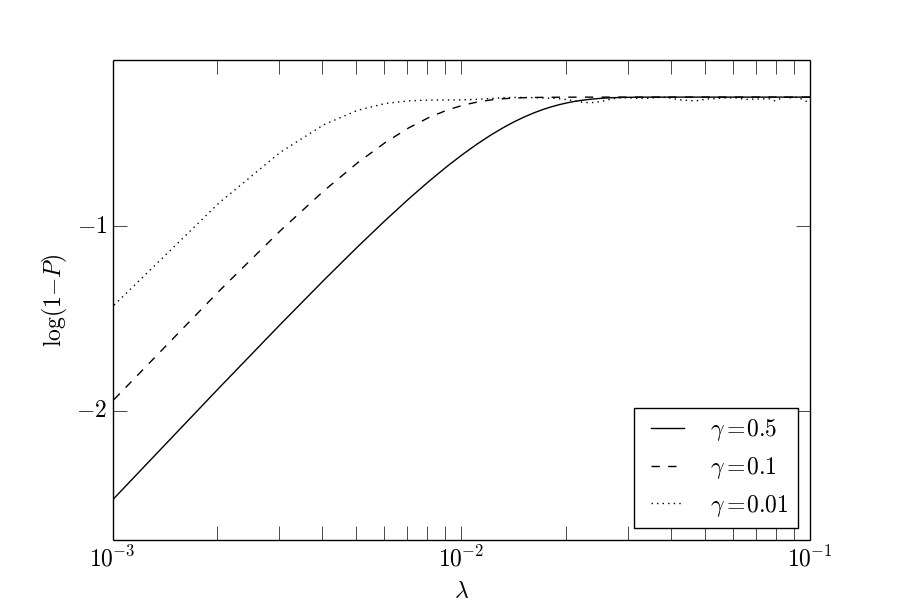}
 \caption{The $\log_{10}(1-P)$ as a function of $\lambda$ for different values of $\gamma$. The results show a straight line
 up to a point where $\lambda$ becomes relatively large. This is in accordance with previous results that showed how the 
 purity decay is proportional to $\lambda^2$ for small $\lambda$.}
 \label{logpurity}
\end{figure} 

The effect indeed must disappear as $\lambda$ reaches identity. The  fact
that the central system does not directly interact with the far environment
becomes irrelevant as any initial state will soon mix the excitations of
the degrees of freedom of the central system with those of the near
environment.

This article evolves around the phenomenon that increasing coupling between
near and far environment slows decoherence.  The case of small couplings is
discussed extensively in the perturbative regime in~\cite{moreno2015,
gorin2015generalized}, while the case or strong absorption is discussed in some
generality in~\cite{Zanardi14}.  The intermediate region, is at this moment
essentially only accessible numerically except for very simple integrable
models~\cite{TorresandSeligman}.  We have given a survey of many options for
the intermediate region, and we have consistently found the effect under
discussion. Yet we have not approached the decoherence free limit.  This may be
related to the finite size we have to use, and which is not implicit in the
rate equation approach used by Campos and Zanardi. We have restricted our
studies to two-body interactions whose convergence for large spaces is known to
be be as $1/\ln{N}$ and thus for numerical purposes inadequate. Another
important point is the restriction to dephasing we used.  We can lift this
restriction by adding the magnetic field kicks to the central system. 
In \fref{fig:internal:field:central}
we show with parameters otherwise the same as in \fref{puritytime}, that this
slows the decoherence much more than dephasing. Thus, while our findings do not
confirm the decoupling in the strong coupling limit between near and far
environment, it certainly does not contradict the decoupling.

\begin{figure} 
 \centering
 \includegraphics[scale=0.5]{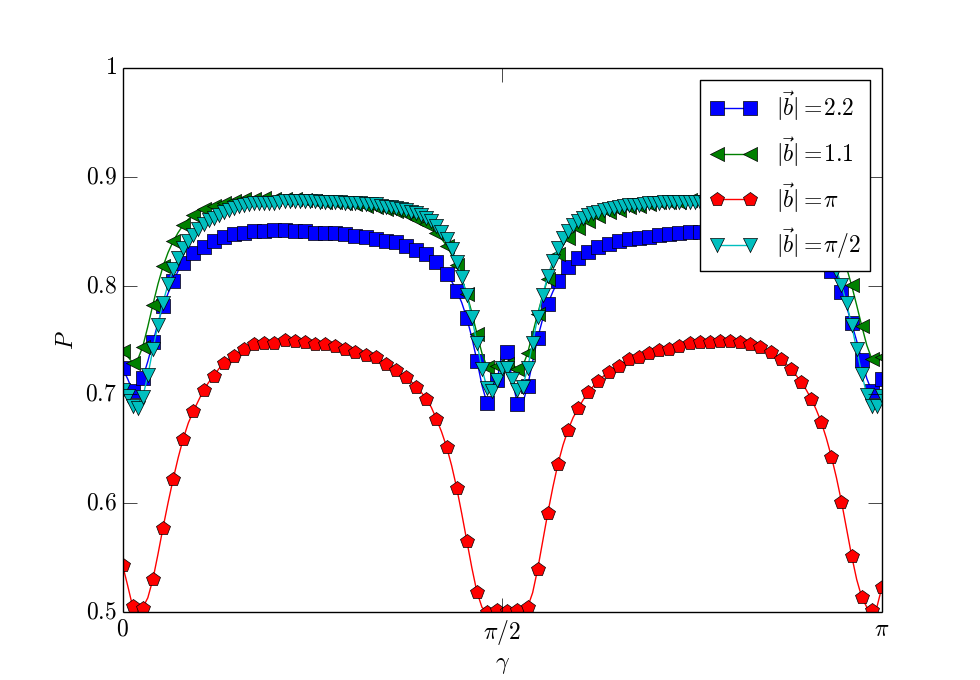}
 \caption{Purity at a fixed time, when adding an internal magnetic field $\vec
 b$ at a $pi/4$ angle with respect to Ising. The effect is preserved, i.e.
 larger couplings imply larger values of purities for the central system, 
 as long as the periodicity in this parameter is not coming to bear. The
 sensitivity on the magnetic field is large and thus the non-dephasing terms
 are important.
 }
 \label{fig:internal:field:central}
\end{figure} 

\section{Two qubits as central system} 
\label{twoqubits}
As we discussed above, a central system composed of two qubits is of great
relevance because it is the building block for universal quantum computation
and other important protocols~\cite{nielsen2010}. Decoherence and the
entanglement shared in a two-qubit system is the subject of many interesting
papers, mainly in the
context of cavity quantum electrodynamics using Markovian master equations
in Lindblad form~\cite{Bosco11,Chaudhry14}.

The aim of this section is to study the evolution of the internal entanglement
in the central system and how it is affected by the presence of a nested
environment. If we have two quits we can in a natural way think of them as
being coupled to the same or different environments and also of
the two couplings being of equal or different strength; in the latter case we
have the option that of one of the two is uncoupled. Such a situation we
call a {\it spectator model}~\cite{pinedaPRA06}. We will focus our
attention in this particular configuration in which one of the qubits plays the
role of an observer. Furthermore, we assume that the two qubits are
non-interacting, avoiding the influence of this internal coupling on the
entanglement evolution in the central system.  We use as entanglement measure
the  concurrence defined in Eq.~\ref{concurrencia}. In~\fref{concurrence} we
show the evolution of the concurrence for an initial Bell pair in a nested
environment for same parameters and configuration of the environments as
in~\fref{puritytime}. We see that the behaviour is indeed quite similar, which
strengthens our point, that this mechanism may actually be appropriate in very
general terms to improve conditions for quantum information processing and
quantum computing.  The phenomenon of entanglement sudden death is
present in all the curves so we can actually use the coupling $\gamma$ in
order to delay it.  The inset plot also shows the evolution of purity over
time for the Bell pair. The results are similar to the case of one qubit
in~\fref{puritytime}. 
 
As in the case of a single qubit, it is interesting to see the evolution of the
concurrence for fixed couplings varying the number of
connections $\nu$ between near and far environment as shown
in \fref{c-connections}.  The connections have been varied in the same manner
as in \fref{puritytime}.  Notice that the scaling of $\gamma$ is still
valid for this configuration, showing the robustness of the effect and
essentially the same behaviour for the evolution of the concurrence. 

\begin{figure} 
 \centering
 \includegraphics[scale=0.45]{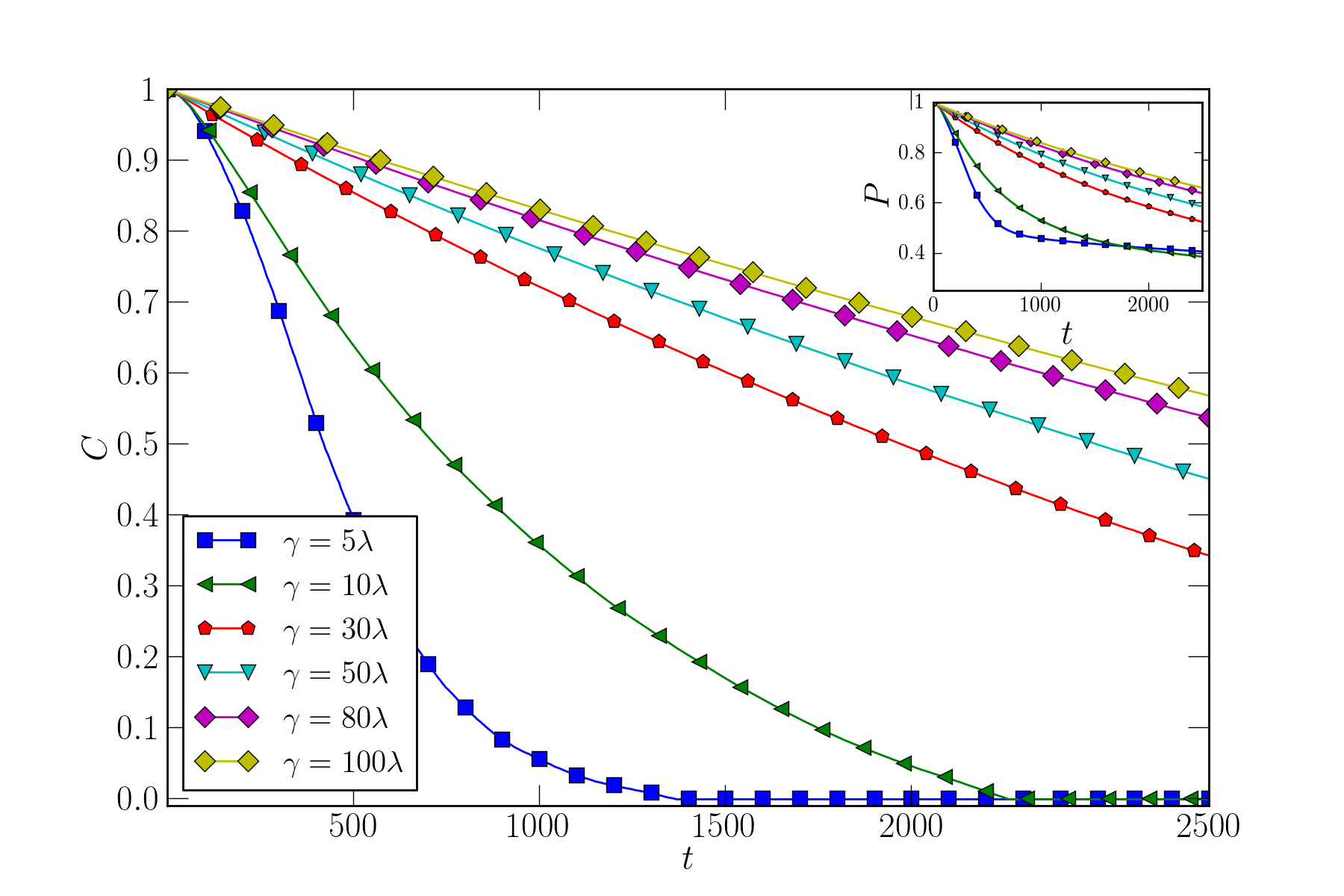}
 \caption{Time evolution of the concurrence (main figure) and purity (inset)
  for different values of $\gamma$ for an initial Bell state
 $(|00\>+|11\>)/\sqrt{2}$ for two-qubits as central system in the {\it
 spectator} configuration. 
We have implemented this by adding a spectator to the configuration of
\fref{puritytime}. We observe a similar behaviour of slowing down of
decoherence as the coupling to the far
environment increases. This effect is also reflected in the behaviour of the
concurrence in the main figure. 
} \label{concurrence}
\end{figure} 
\begin{figure} 
 \centering
 \includegraphics[scale=0.45]{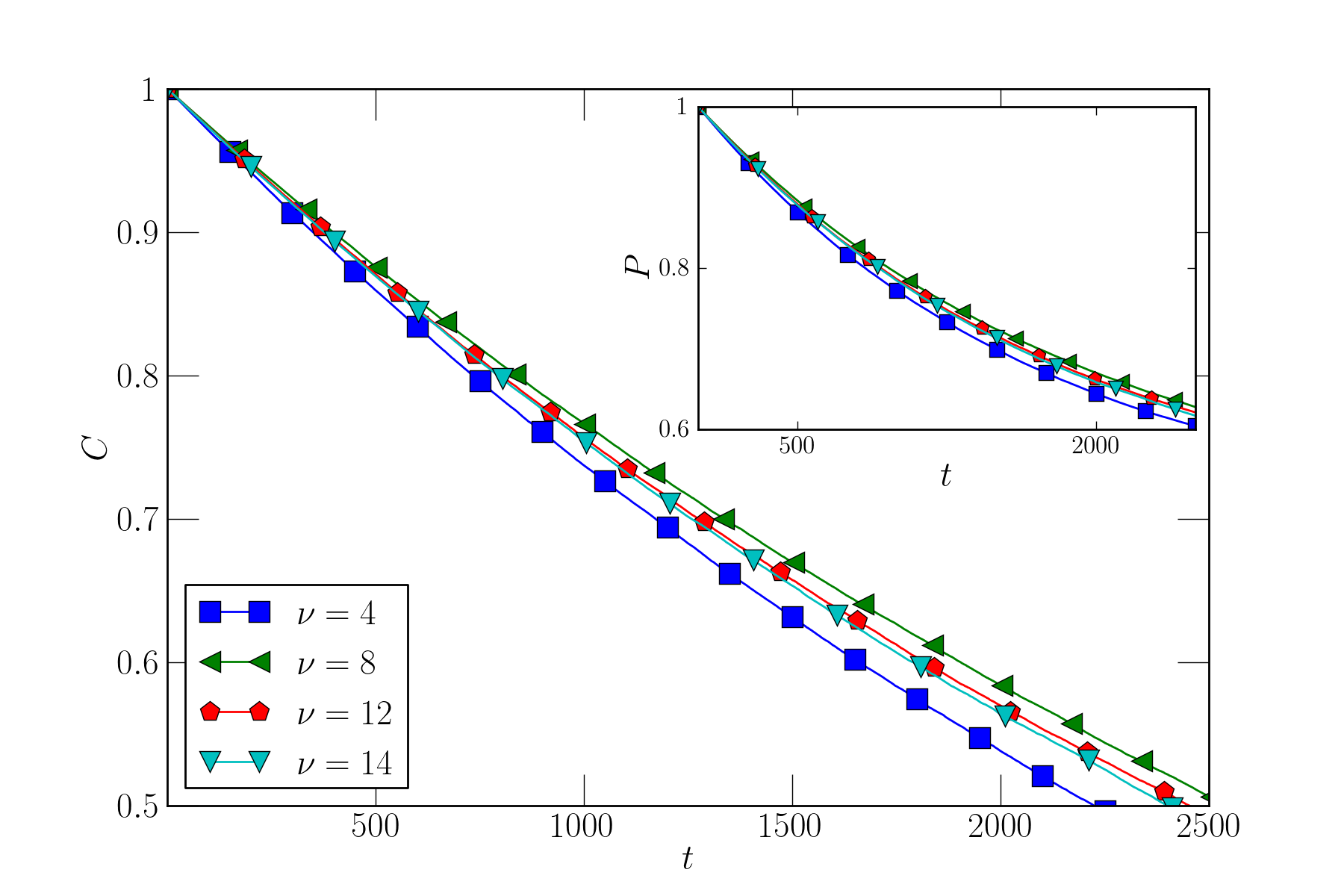}
 \caption{Time evolution of the concurrence (main figure) and purity (inset) for an initial Bell state $(|00\>+|11\>)/\sqrt{2}$ for two-qubits as central system 
 in {\it spectator} configuration varying the number of connections $\nu$ between environments (near and far). 
 In each case the rescaled parameter $\gamma^{\prime}=\gamma/\sqrt{\nu}$ is used. We observe that the scaling observed 
 in~\fref{gammascaled} is also approximately valid for the case of a two-qubit system.}

 \label{c-connections}
\end{figure} 
\subsection{$C$-$P$ diagram} 
A useful tool to characterize the decoherence process of a two-qubit system is
the so called concurrence-purity diagram or $C$-$P$ plane~\cite{ziman:052325}.
One point on this diagram gives the value of mixedness and entanglement 
simultaneously. Those quantum states that for a definite
value of the purity can reach the maximum degree of entanglement are known as
maximally entangled mixed states~\cite{Izaka00}.  Looking at the behaviour on the $C$-$P$ plane of an
initial pure maximally entangled state under a local quantum channel (which is
the case of the spectator configuration) we can in principle characterize the
corresponding quantum process suffered by the system~\cite{ziman:052325}.
Under the action of unital channels, Bell states are mapped to the region
between the corresponding to Werner states and the one corresponding to 
the action of dephasing channels, over Bell states. 
It is
worthwhile to explore the behaviour of our system in this 
plane, when an internal magnetic field is applied (otherwise, we would
simply have dephasing, and thus we would lie in the lower curve). 

In~\fref{cp1} we show a set of typical $C$-$P$ diagrams with fixed  $\lambda$
and varying $\gamma$.  The first observation is that the quantum channel
induced by the KI for the parameters we have used is actually unital, all the
curves are in the region of unital channels.  The top figure shows that the
lines tend to follow the Werner state behaviour quite closely as we increase
the strength of the interaction between near and far environment (though we
should remember, that actually they might not be Werner states).
From the bottom figure is
clear that for a sufficiently large number of connections between environments, the
increasing of $\gamma$ is no longer effective to improve the coherence in the
central system.  The saturation is reached faster when there is enough
connectivity of the environments.  

\begin{figure} 
 \centering
 \includegraphics[scale=0.45,keepaspectratio=true]{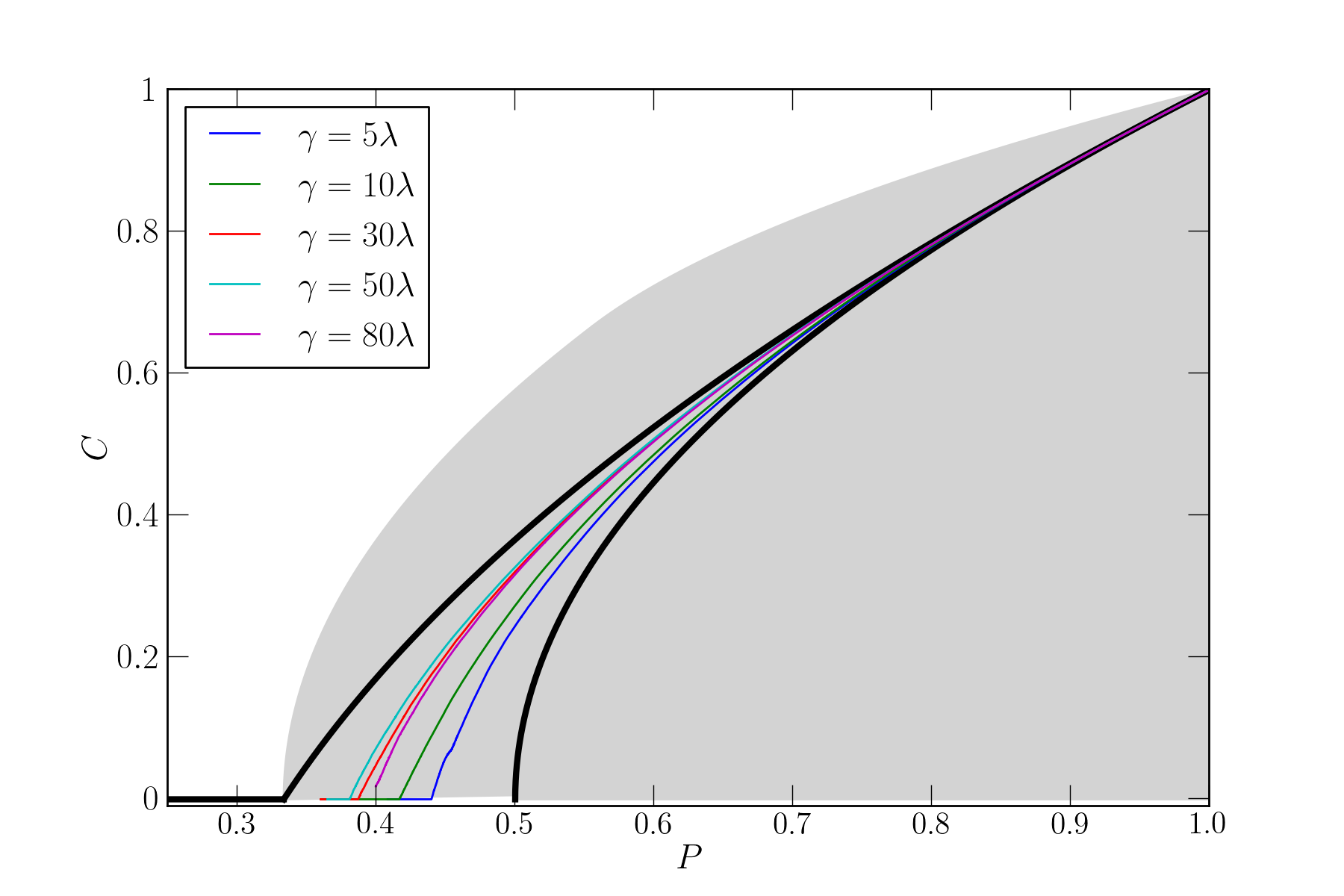}\\
 \includegraphics[scale=0.45,keepaspectratio=true]{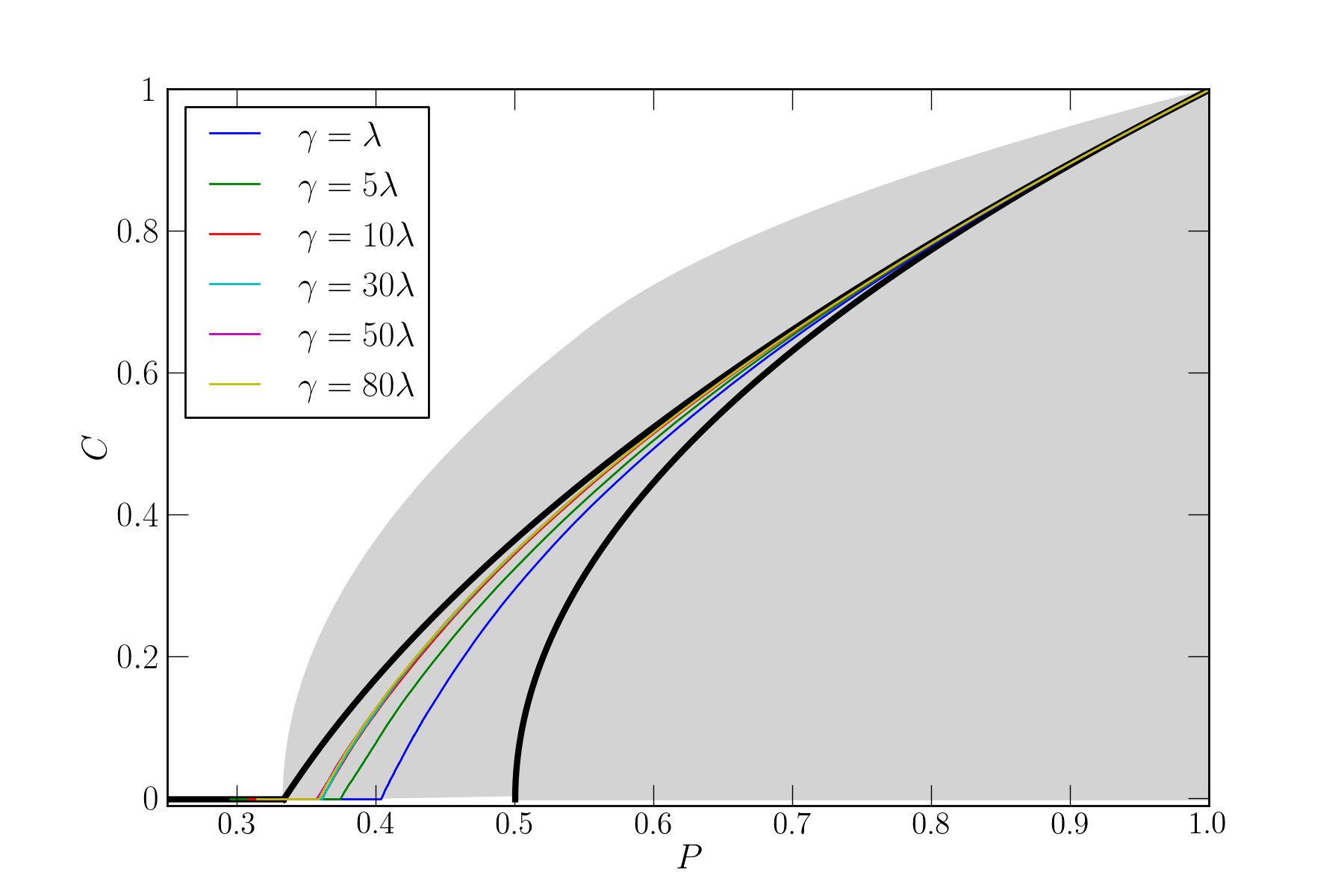}
 \caption{$C$-$P$ diagrams for different values of $\gamma$ and $t\in [0,4000]$
 for an initial Bell state.  The light grey area shows the region of all
 physical states of two qubits, bounded by above by the maximally entangled
 mixed states.  The  region bounded by the curves  for Werner states and
 for Bell states under phase damping channels (thick lines) define the image of
 a Bell pair under the set of local unital operations~\cite{ziman:052325}.
 The top figure corresponds to the configuration illustrated in~\fref{puritytime}.
The bottom figure corresponds  to a configuration with $\nu=16$ connections
 between environments (near and far). Parameters are the same as in~\ref{cp1}.
 We observe that increasing the connectivity diminishes the dispersion of purity 
 for a fixed value of concurrence.
 }
 \label{cp1}
\end{figure} 

\section{Conclusions} 
\label{sec:conclusions}
We have explored numerically various aspects of the effect of nested
environments on a central system using the kicked Ising chain model taking
advantage its map structure, which allows simple calculations, while still
being a many-body system.  Our departing point was the growing evidence, that
for a situation with a central system coupled weakly to a near environment and
no (or negligible) direct coupling to a far environment increasing the coupling
between near and far environment slows decoherence in the near environment.
This effect was confirmed over a wide range of situations for a central system
with a dephasing or a more general coupling to the near environment as well as
for a two qubit system in a Bell state with one of the qubits being in a
(non-interacting) spectator situation. We demonstrate a similar behaviour for
the concurrence, which is an essential point for the usefulness of the
encountered effect in the context of quantum information.
\ack 
We thank Thomas Gorin and H\'ector Moreno for very stimulating discussions.
Support by the  projects CONACyT 153190, CONACyT 219993,  UNAM-PAPIIT IN111015 and
UNAM-PAPIIT IG101113 are acknowledged.
One of us (CGG) is grateful to CONACyT for financial support
under the doctoral fellowship No. 385108.

\section*{References}
\bibliographystyle{unsrt} 
\bibliography{mis_referencias}
\end{document}